\documentstyle[prl,aps]{revtex}
\input epsf.sty

\newcommand{\be}{\begin{equation}}
\newcommand{\ee}{\end{equation}}
\newcommand{\bd}{\begin{displaymath}}
\newcommand{\ed}{\end{displaymath}}
\newcommand{\ba}{\begin{array}}
\newcommand{\ea}{\end{array}}
\newcommand{\bt}{\begin{tabular}}
\newcommand{\et}{\end{tabular}}
\newcommand{\bc}{\begin{center}}
\newcommand{\ec}{\end{center}}
\newcommand{\bn}{\begin{enumerate}}
\newcommand{\en}{\end{enumerate}}
\newcommand{\bi}{\begin{itemize}}
\newcommand{\ei}{\end{itemize}}
\newcommand{\bqr}{\begin{eqnarray}}
\newcommand{\eqr}{\end{eqnarray}}
\newcommand{\bfig}{\begin{figure}[tbp]}
\newcommand{\efig}{\end{figure}}
\newcommand{\btab}{\begin{table}[ht]}
\newcommand{\etab}{\end{tabular}\ec\end{table}}
\newcommand{\bl}{\begin{large}}
\newcommand{\el}{\end{large}}

\newcommand{\nb}{\nonumber}

\begin{document}
\draft

\twocolumn[\columnwidth\textwidth\csname@twocolumnfalse\endcsname

\title{Odd-even mass differences in self-consistent Mean-Field calculations.}
\author{
T. Duguet,$^{1}$ 
P. Bonche,$^{1}$
P.-H. Heenen,$^{2}$ and
J. Meyer,$^{3}$
}
\address{$^1$Service de Physique Th\'eorique, CEA Saclay, 91191 Gif sur Yvette Cedex, France}
\address{$^2$Service de Physique Nucl\'eaire Th\'eorique, Universit\'e Libre de Bruxelles, C.P 229, B-1050 Bruxelles, Belgium}
\address{$^3$IPN de Lyon, CNRS-IN2P3 / Universit\'e Claude Bernard Lyon 1, 43, Bd. du 11.11.18, 69622 Villeurbanne Cedex, France}
\date{\today}

\maketitle
 
\begin{abstract}

The odd-even mass staggering (OES) of nuclei is analyzed in the context of self-consistent mean-field calculations. The procedure developed allows to understand the OES for spherical as well as for deformed nuclei. Comparison with results at the Hartree-Fock level shows the non-perturbative effect on this observable of the inclusion of pairing correlations.

\addvspace{5mm}

\end{abstract}]

\narrowtext


Odd-even staggering of binding energies is a common phenomenon of several finite many-fermion systems. In nuclei, it has been attributed to an evidence of pairing correlations\cite{Boh8}. Assuming that masses are smooth functions of the number of neutrons and protons except for pairing effects, simple expressions have been derived for the gap parameter $\Delta$ based on the differences between binding energies of even and odd nuclei\cite{BM75,jensen}. Detailed analyses\cite{mad} and pairing adjustments\cite{mol} have been based on these expressions. The simplest example is the well-known three-point mass formula: 

\begin{equation}
\Delta^{(3)}(N) = \frac{(-1)^{N}}{2} \, [E(N\!+\!1)-2E(N)+E(N\!-\!1)] \,  \\  \vspace{0.3cm}
\label{defd3}
\end{equation}
where N (Z) is the number of neutrons (protons).
However, a study of the OES in light alkali-metal clusters and light N=Z nuclei~\cite{Haa98} has lead to the conclusion that this phenomenon was not due to pairing correlations but rather to deformation effects. That work motivated a study by Satula et al.~\cite{SDN98} about the ``mean-field'' contribution to the OES in nuclei, especially from deformation. To isolate mean-field effects, they set the pairing force to zero and performed Hartree-Fock (HF) calculations for light deformed nuclei. At this level, they already discovered an odd-even staggering, itself oscillating: 

\begin{eqnarray}
&\begin{array}{lccc}
\Delta^{(3)}_{HF}(N) &\approx& 0 & \, {\rm if} \ N \ {\rm is \ odd} \, \,  , \\
 & \approx & \frac{e_{k}-e_{k-1}}{2} & \, {\rm if} \ N \ {\rm is \ even}  \, \,  ,
\end{array}&
\label{delta3hf}
\end{eqnarray}

\noindent where $(e_{k}-e_{k-1})$ stands for the possible gap around the Fermi level and is zero in spherical nuclei (apart for magic numbers) but different from zero for deformed nuclei because of the spread doubly-degenerated spectrum. Thus, they found the deformation (Jahn-Teller effect) to be responsible for a direct contribution to the three point odd-even mass formula.  Consequently, in presence of pairing correlations $\Delta^{(3)}_{odd}$ could be a measure of pairing effect only,  whereas $\Delta^{(3)}_{even}$ will contain an additional contribution responsible for the staggering of this quantity. Note that, such a scheme cannot account for the same oscillation in spherical nuclei. 

The aim of the present study is to analyze for all nuclei the contributions to odd-even mass differences in a fully self-consistent mean-field picture and to extract a quantity more directly related to pairing correlations. After the presentation of the results, our approach will be compared to that of the Ref.~\cite{SDN98}.



Finite-difference mass formulas are intended to extract the quickly varying part of the energy as a function of the neutron or proton number. The underlying assumption is the possible division of the energy into a smooth and a rapidly varying part. OES can be viewed as a probe of the difference of structure between odd and even nuclei, due to the polarization effect by the odd nucleon and to the reduction of pairing by its blocking.

The energy would be smooth if those differences of structure would not exist. Thus the smooth part can be defined as E$^{HFBE}$ (HFBE meaning Hartree-Fock-Bogolyubov {\it Even}) which is the energy obtained when all nuclei are calculated as if they were even ones (no blocking and no breaking of the time reversal symmetry in odd nuclei). This definition of E$^{HFBE}$ has already been used in~\cite{bend} in a similar context. The theoretical energy of an odd nucleus can be decomposed as: 

\begin{eqnarray}
E^{HFB}(N) &=& E^{HFBE}(N) + [E^{HFB}(N) - E^{HFBE}(N)] \nb  \\
&=& E^{HFBE}(N) + \overbrace{E^{pol}(N) + \Delta(N)}  \label{defener} \, ,
\end{eqnarray}
where $E^{pol}(N)$ is the difference of binding energy due to the polarization of the core by the odd nucleon, and $\Delta(N)$ is the positive contribution specifically related to the blocking effect. 

From the explicit calculation of both $E^{HFB}$ and $E^{HFBE}$, one obtains $E^{pol}(N) + \Delta(N)$ through the odd-even mass formulas which are quantities comparable to experiment. The separation between $E^{pol}$ and $\Delta(N)$ is not discussed in this paper.


Following eq. \ref{defener}, the three point mass formula becomes:
\begin{equation}
\Delta^{(3)}_{HFB}(N) =  \Delta^{(3)}_{HFBE}(N) + \Delta^{(3)}_{pairing + pol}(N)  \, \, . \vspace{0.2cm}
\label{d3}
\end{equation}
As E$^{HFBE}$ varies smoothly with N, one can write:
\be
\Delta^{(3)}_{HFBE}(N) \approx  \frac{(-1)^{N}}{2} \, \left. \frac{\partial^{2} E^{HFBE}}{\partial N^{2}} \, \right|_{N} \,  \, .
\label{d3approx}
\ee
We also obtain

\begin{eqnarray}
&\begin{array}{rclcl}
\Delta^{(3)}_{pairing}(N) &=&  \Delta(N)  &\qquad&{\rm if} \ N \ {\rm is \ odd} \nb \\ [8pt]  
&=& \frac{\Delta(N\!-\!1) +\Delta(N\!+\!1)}{2}  & &{\rm if} \ N \ {\rm is \ even,}  \nb
\end{array}&  \nb
\end{eqnarray}

\noindent and similarly for $\Delta^{(3)}_{pol}(N)$. Similar expressions can be written for higher order formulas. We will use below the five-point formula (fourth-order):

\begin{equation}
\Delta^{(5)}_{HFB}(N) = \Delta^{(5)}_{HFBE}(N) + \Delta^{(5)}_{pairing + pol}(N)  \, .
\label{d5} 
\end{equation}


We have calculated energies using the formalism and forces detailed in Ref.~\cite{tera,rigol}. It is based on the  self-consistent  HFB method with an approximate particle number projection using the Lipkin-Nogami prescription. It should be mentioned that the time-odd components of the force are included since the time-reversal symmetry is explicitly broken in the calculation of odd nuclei.



To avoid the contributions from deformation we first study the odd-even mass staggering along an isotopic chain of spherical nuclei. Seventy ground-states are calculated along the tin isotopic chain, from $^{100}$Sn to $^{169}$Sn. Each odd-N nucleus is calculated twice: first, with the fully self-consistent HFB scheme (several 1 quasi-particle (qp) configurations are investigated to get the lowest in energy); and second, as a HFB vacuum requiring only an odd average number of neutrons (HFBE case).

\begin{figure}
\begin{center}
\leavevmode
\epsfxsize=6.5cm
\epsfbox{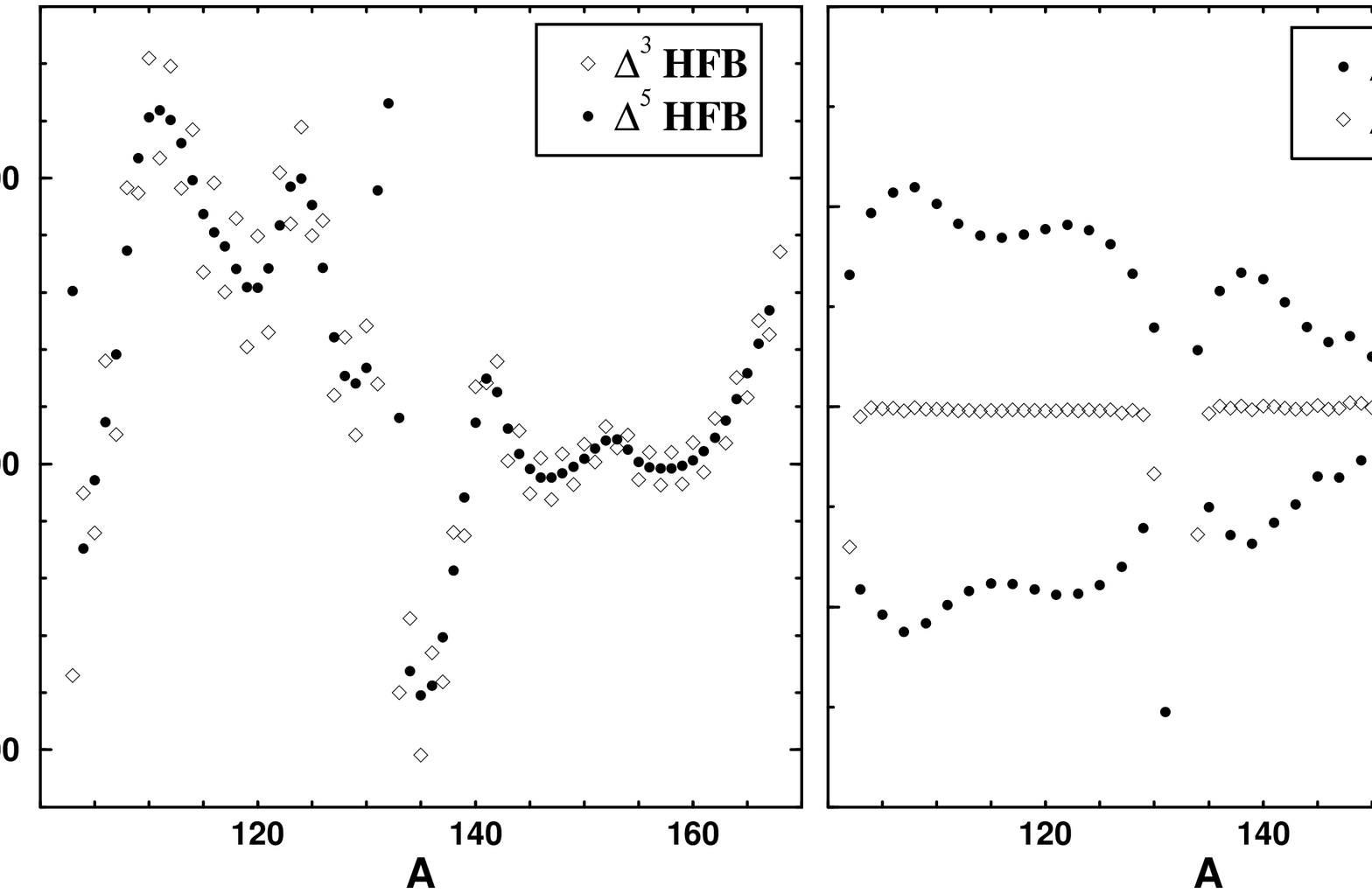}
\end{center}
\caption{Left: calculated odd-even mass differences $\Delta^{(3)}_{HFB}$ and $\Delta^{(5)}_{HFB}$ for the tin isotopic line from $^{100}$Sn to $^{169}$Sn. Right: $\Delta^{(3)}_{HFBE}$ and $\Delta^{(5)}_{HFBE}$ for the same nuclei.}
\label{deltashfbSn}
\end{figure}

Calculated $\Delta^{(3)}_{HFB}(N)$ and $\Delta^{(5)}_{HFB}(N)$ along this chain are given on the left panel of Fig.~\ref{deltashfbSn}. A staggering is observed for $\Delta^{(3)}(N)$ whereas no staggering occurs for $\Delta^{(5)}(N)$.

To understand these observations, let us focus first on the right panel of Fig.~\ref{deltashfbSn} which displays the contributions of the smooth part of the energy E$^{HFBE}$ to $\Delta^{(3)}$ and $\Delta^{(5)}$ (eq. \ref{d3approx} and \ref{d5}). Apart for the magic number $N=82$, this figure shows that $\Delta^{(3)}_{HFBE} \gg  \Delta^{(5)}_{HFBE} \simeq 0$. This legitimates the identification of E$^{HFBE}$ as the smooth part of the energy. Indeed, the larger the order of the formula, the smaller the expected contribution coming from an existing smooth part of the energy. 

Assuming that $E^{pol}(N) + \Delta(N)$ does not vary significantly over a few consecutive odd nuclei and gives similar contributions to $\Delta^{(3)}$ and $\Delta^{(5)}$, one can write:

\begin{eqnarray}
\Delta^{(3)}_{HFB}(N) -  \Delta^{(5)}_{HFB}(N) &\approx& \Delta^{(3)}_{HFBE}(N)  \label{staggd3} \\
\Delta^{(5)}_{HFB}(N) &\approx& \Delta^{(5)}_{pairing+pol}(N)  \nb  \, .
\end{eqnarray}

\begin{figure}
\begin{center}
\leavevmode
\epsfxsize=6.5cm
\epsfbox{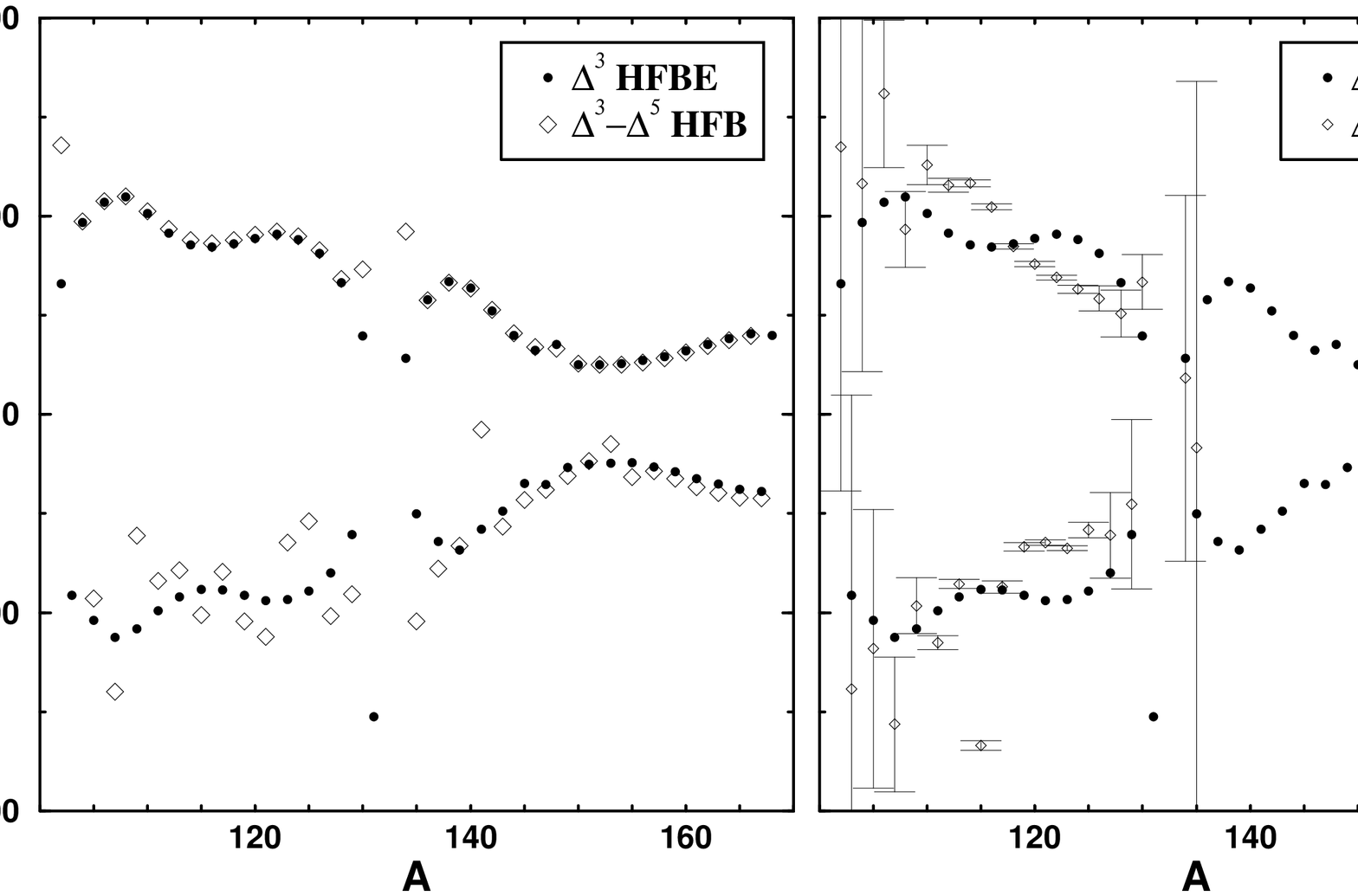}
\end{center}
\caption{Left: $\Delta^{(3)}_{HFBE}$ is compared to $\Delta^{(3)}_{HFB}$ - $\Delta^{(5)}_{HFB}$ along the tin isotopic line. Right: comparison of $\Delta^{(3)}_{HFBE}$ with experiment (see text).}
\label{d3-d5Sn}
\end{figure}

To justify this equation, we compare on the left panel of Fig.~\ref{d3-d5Sn} both sides of eq.~\ref{staggd3}. The agreement is impressive along the whole isotopic line (except for the magic number $N = 82$). Some one-qp configurations for odd nuclei are hard to converge leading to slight irregularities in the curve $\Delta^{(3)}-\Delta^{(5)}$ HFB. One can notice that $\Delta^{(3)}_{HFBE}(N)$ is equal, in absolute value, for odd and even neighbors. The right panel of Fig.~\ref{d3-d5Sn} provides a comparison between $\Delta^{(3)}_{HFBE}$ and $\Delta^{(3)}_{Exp}(N)-\Delta^{(5)}_{Exp}(N)$ deduced from experimentally known masses~\cite{audi}. The agreement between the staggering of $\Delta^{(3)}_{Exp}$ around $\Delta^{(5)}_{Exp}$ and $\Delta^{(3)}_{HFBE}$ is very good as well.

Eq.~\ref{staggd3} and the above numerical results shed a new light on the OES problem. Indeed, the $\Delta^{(3)}$ staggering must be understood as an oscillation around $\Delta^{(5)}$ due to the contribution to $\Delta^{(3)}$ of the smooth part of the energy. Moreover, $\Delta^{(5)}$ is found to be a good measure of the rapidly varying part of the energy $\Delta^{(5)}_{pairing+pol}(N)$ which we want to extract while $\Delta^{(3)}_{odd}$ still contains smooth contributions. Finally, these results fully justify the separation of the energy performed in eq.~\ref{defener}. We want to make clear that although our calculated $\Delta^{(3)}_{HFB}(N)$ overestimates experiment (not shown here), the oscillation of $\Delta^{(3)}_{HFB}(N)$ is in agreement with experiment. This underlines the decoupling between these two aspects.


To extend the above analysis to deformed nuclei, we have calculated all cerium isotopes from $^{118}$Ce to $^{166}$Ce. The ground-state quadrupole deformation along this isotopic line undergoes large variations. It goes from a region of strong prolate deformation around  $^{118}$Ce ($\beta_{2} \approx 0.37$), through the spherical $^{140}$Ce nucleus, to another prolate deformation region ($\beta_{2} \approx 0.31$ around $^{160}$Ce).

\begin{figure}
\begin{center}
\leavevmode
\epsfxsize=6.5cm
\epsfbox{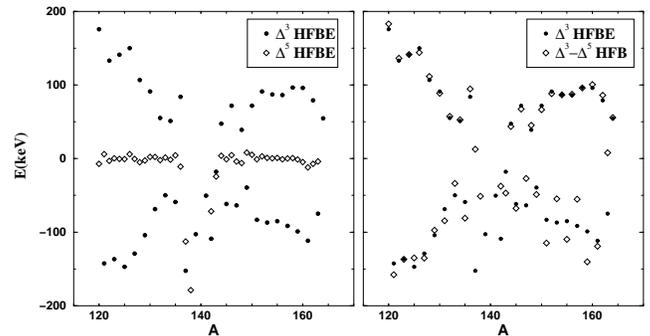}
\end{center}
\caption{Left: same as the right panel of Fig.~\ref{deltashfbSn} for cerium isotopes. Right: same as the left panel of Fig.~\ref{d3-d5Sn} for cerium isotopes.}
\label{deltashfbeCe}
\end{figure}

On the left panel of Fig.~\ref{deltashfbeCe} are shown $\Delta^{(3)}_{HFBE}$ and $\Delta^{(5)}_{HFBE}$. The comparison between $\Delta^{(3)}_{HFB}-\Delta^{(5)}_{HFB}$ and $\Delta^{(3)}_{HFBE}$ is presented on the right panel. The same kind of results and agreements as for the tin isotopes is obtained: $\Delta^{(3)}_{HFBE} \gg  \Delta^{(5)}_{HFBE} \approx 0$ and $\Delta^{(3)}_{HFB}(N)-\Delta^{(5)}_{HFB}(N) \approx \Delta^{(3)}_{HFBE}(N)$. The oscillations of $\Delta^{(3)}$ around $\Delta^{(5)}$ are once again well reproduced by the contribution coming from $E^{HFBE}$. Comparison with experiment has not been included because the error bars on the cerium masses are too large to make it relevant. These calculations show that deformation does not modify the conclusions drawn above for spherical nuclei.

In our separation procedure, the deformation manifests itself in several ways. First, the main effect of deformation is present in E$^{HFBE}$, apart from regions of shape transition and shape coexistence where the deformation varies suddenly with N.  The smooth contribution of the deformation to the OES is thus extracted through $\Delta^{(n)}_{HFBE}$. There is an additional contribution related to odd-even effect. This term is usually small~\cite{dug} and manifests itself in E$^{pol}$. As seen from our numerical results and the above explanation, the separation given by eq.~\ref{defener} is well defined whether the nuclei are deformed or not. Deformation will also influence the pairing part of the energy through correlations between particle-hole and particle-particle channels. The evolution of $\Delta^{(n)}_{pairing}$ with deformation might give some informations about the behavior of the pairing force with respect to this degree of freedom.

\begin{figure}
\begin{center}
\leavevmode
\epsfxsize=6.1cm
\epsfbox{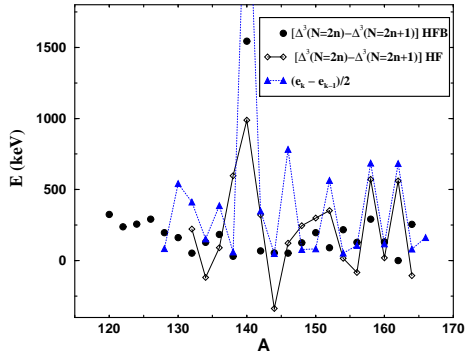}
\end{center}
\caption{Comparison of the staggering of $\Delta^{(3)}$ for the cerium isotopes in HFB and Hartree-Fock calculations. The simple expression $(e_{k}-e_{k-1})/2$ is also given.}
\label{compHFHFB}
\end{figure}

As it has been suggested to use HF energies as a reference to study the $\Delta^{(3)}$ staggering~\cite{SDN98}, we compare in Fig.~\ref{compHFHFB} the staggering $\Delta^{(3)}(2n) - \Delta^{(3)}(2n+1)$ obtained with full HFB calculations and Hartree-Fock ones (without time-reversal symmetry breaking, see~\cite{SDN98}) for the cerium isotopic chain. The single-particle splitting around the Fermi level (eigen-states of the HF field) $(e_{k}-e_{k-1})/2$ in even-N nuclei, due to the deformation, is also shown as it is presumably related to this staggering (eq.~\ref{delta3hf}). The HF calculations do not reproduce the staggering of $\Delta^{(3)}_{HFB}$: the global trend is different between the two cases and for most nuclei the difference between the two calculations exceeds 200 keV. One can also see that the difference between $\Delta^{(3)}_{HFB}(N=2n)$ and $\Delta^{(3)}_{HFB}(N=2n+1)$ is not a measure of the splitting of the single-particle levels $(e_{k}-e_{k-1})/2$ around the Fermi energy. The latter conclusion for deformed nuclei holds of course also for spherical ones.

In summary, we have proposed a new analysis for the odd-even mass staggering based on self-consistent mean-field calculations. It assumes the definition of a ``virtual'' odd nucleus, having the structure of an even one, as the underlying structure of the ``real'' nucleus. The notion of underlying structure is justified by the validation of this energy separation obtained for a large number of odd nuclei, whatever the type of qp (shell, spin, parity) blocked self-consistently on this vacuum to describe the odd nuclei. Instead of being a single-particle excitation on an even core, an odd nucleus is seen as a qp excitation constructed on an odd fully paired core. It has permitted to identify the staggering of $\Delta^{(3)}$ for spherical, as well as for deformed nuclei, as an oscillation around $\Delta^{(5)}$ due to the contribution of this fully paired core. Moreover, $\Delta^{(5)}$ has been found to be the right quantity to extract the pairing part of the odd-even staggering. This conclusion has already been drawn from numerical comparison of HFBCS calculations with experiment for several possible theoretical quantities measuring the gap parameter~\cite{bend}.

Finally, our results do not confirm the ones obtained at the HF level that the staggering of $\Delta^{(3)}$ is a measure of the single-particle splitting around the Fermi energy in deformed nuclei. A forthcoming article will present a full analysis of the above and complementary results.

Methods~\cite{valor,rod} have been recently introduced in which the correlations beyond mean field due to symmetry restorations are included. Such developments require a reanalysis of the different contributions to the EOS. It is one of the future steps of our study within the above framework.


\end{document}